# TWO-WAY ↔ QUANTUM COMMUNICATION USING FOUR-QUBIT CLUSTER STATE: MUTUAL EXCHANGE OF QUANTUM INFORMATION

VIKRAM VERMA

*Department of Physics, Deshbandhu College, University of Delhi,*
*New Delhi-110019, India.*
*vikramverma18@gmail.com*

MITALI SISODIA

*Department of Physics, Indian Institute of Technology Jodhpur,*
*Jodhpur, Rajasthan 342011, India.*
**Current address-** *Department of Physics, Indian Institute of Technology Delhi,*
*New Delhi, Delhi 110016 India.*
*mitalisisodiyadc@gmail.com*

In the present study, we have proposed a scheme for two-way quantum communication in which the two legitimate participants mutually exchange their quantum information to each other by using a four-qubit cluster state as the quantum channel. Recently, by utilizing four-qubit cluster state as the quantum channel, Kazemikhah et al. [*Int. J. Theor. Phys.,* **60** (2021) 378] tried to design a scheme for mutual exchange of quantum information between two legitimate participants. However, in the present study it has been shown that in their scheme the transmission of quantum information cannot be realized because the two participants are not entangled to each other due to a trivial conceptual mistake made by Kazemikhah et al. in the description of the quantum channel. Here, we have shown that two legitimate participants can teleport quantum information states to each other by using four-qubit cluster state as the quantum channel, provided they co-operate with each other and perform non local controlled phase gate operation. If both participants do not co-operate with each other, then no one can reconstruct the information sent to them, and therefore the exchange of information is possible only when both participants are honest to each other.

## 1. Introduction

Quantum teleportation (QT), an important aspect of quantum entanglement, allows the transmission of an unknown quantum state by using a previously shared entanglement and classical communication. The first idea of QT was designed in 1993 by Bennett et al.[1] for a single-qubit state by using a Bell state. Subsequently, various schemes have been proposed for the teleportation of multi-qubit states using various types of entangled states, like EPR state,[2] GHZ state,[3,4] GHZ-like state,[5,6] W state,[7,8] W-like state,[9] cluster state[10-12] et al. In 2017, Mitali et al.[13] proposed a generalized scheme of quantum teleportation using optimized amount of quantum resource that means any unknown quantum state of the form of $|\psi\rangle = \sum_{i=1}^{m} \alpha_i |x_i\rangle$ can be teleported with the help of some of Bell states which is the minimum number of entangled qubits.

As the expansion of QT, many variants of QT schemes, such as quantum information splitting (QIS),[14,15] quantum secret sharing (QSS),[16,17] controlled quantum teleportation (CQT)[18,19] and bidirectional and bidirectional controlled quantum teleportation (BQT and BCQT)[20-34] etc. have been proposed.

Among these schemes of QT, BQT which is a two-way quantum communication has attracted more attention because two legitimate users simultaneously exchange their quantum information by using an appropriate entangled quantum channel and classical communication. In 2013, Zha et al.[27] proposed a bidirectional quantum teleportation scheme via five-qubit cluster state. After that, many schemes for BQT have been proposed [20-34] by using various type of quantum resources. For example, a six-qubit cluster state is used in Zhou et al.'s BQT scheme[23] in which Alice can transmit an unknown three-qubit entangled state to the Bob, and at the same time, Bob can transmit an arbitrary single-qubit state to the Alice. In another BQT scheme,[31] a four-qubit cluster state is used by Zhou et al. to teleport bi-directionally two-qubit unknown quantum state. Later, by using eight-qubit entangled state as a quantum channel, Houshmand et al. proposed a new BQT scheme[32] for arbitrary two-qubit unknown quantum states. Quantum resource used in Zhou et al.'s BQT scheme[31] is not entangled, this trivial conceptual mistake is found by the author in Ref. 34.

Recently Kazemikhah et al.[33] have proposed a scheme for BQT using four-qubit cluster state as a quantum channel, but we observed that in their scheme, two legitimate users are not entangled to each other due to trivial mistake in the description of quantum channel. We know that the entanglement between sender and receiver is a necessary ingredient to perform the quantum teleportation scheme successfully and it is not possible to teleport perfectly any unknown quantum state without entanglement. To consider this point, in this paper, following the idea of using non-local operations in designing the schemes for BQT[28-30] by using five and more qubits entangled states as the quantum channel, we propose a new scheme for BQT by using four-qubit cluster state as the quantum channel. A four-qubit cluster state[35] is a highly entangled state of four-qubits and is given by $|\phi_c\rangle_{1234} = \frac{1}{2}(|0000\rangle + |0011\rangle + |1100\rangle - |1111\rangle)_{1234}$. Such cluster state has immense use in one way quantum communication[36-38] and therefore investigating its appropriate use in two way quantum communication is also remarkable. In contrast to the previous communication schemes[36-38] in which the transmission of information is in one direction only, in our propose communication scheme the transmission of information is possible in both directions simultaneously.

The outline of this paper is structured as follows. Section 2 involves two subsections for two-way quantum communication between two legitimate participants Alice and Bob. In Subsection 2.1, we have revised the BQT scheme for arbitrary single-qubit unknown quantum states by using four-qubit cluster state as the quantum channel. In Subsection 2.2, the proposed scheme has been generalized to BQT of multi-qubit states. A comparison has been shown in Section 3. Finally, a brief conclusion is drawn in Section 4.

## 2. Two-way Quantum Communication using Four-Qubit Cluster State as the Quantum Channel

In this section, by using four-qubit cluster state as a resource state, we propose the BQT schemes for arbitrary single-qubit and certain class of multi-qubit unknown quantum states. Here, the two legitimate participants Alice and Bob mutually exchange their unknown quantum states with each other with the help of some non-local controlled phase gate operations. In contrast to the previous BQT schemes[28-30] that also require the use of nonlocal operations, our proposed BQT scheme requires less quantum resource consumption and hence it has comparatively higher intrinsic efficiency[33].

### 2.1. *Mutual exchange of single-qubit information states*

Let us considered two legitimate users Alice and Bob who have arbitrary single-qubit unknown quantum states $|\psi\rangle$ and $|\phi\rangle$ respectively. These unknown quantum states $|\psi\rangle$ and $|\phi\rangle$ are respectively given by

$$|\psi\rangle_a = (a_0|0\rangle + a_1|1\rangle)_a \qquad (1)$$

$$|\phi\rangle_b = (b_0|0\rangle + b_1|1\rangle)_b, \qquad (2)$$

where coefficients $a_0, a_1, b_0\ \&\ b_1$ are unknown and are satisfying the normalization conditions $|a_0|^2 + |a_1|^2 = 1\ \&\ |b_0|^2 + |b_1|^2 = 1$.

Suppose Alice wants to transmit the unknown quantum state $|\psi\rangle$ to Bob and at the same time Bob wants to transmit the unknown quantum state $|\phi\rangle$ to Alice. For this purpose Alice and Bob share a four-qubit cluster state as the quantum channel. The four-qubit cluster state[35] is given by

$$|Q\rangle_{A_1 B_1 A_2 B_2} = \frac{1}{2}(|0000\rangle + |0011\rangle + |1100\rangle - |1111\rangle)_{A_1 B_1 A_2 B_2} \qquad (3)$$

in which qubits pairs ($A_1, A_2$) and ($B_1, B_2$) belong to Alice and Bob respectively.

The combined state of the qubits $a, b, A_1, B_1, A_2, B_2$ is

$$\begin{aligned}
|\chi\rangle_{a b A_1 B_1 A_2 B_2} &= |\psi\rangle_a \otimes |\phi\rangle_b \otimes |Q\rangle_{A_1 B_1 A_2 B_2} \\
&= \tfrac{1}{4}[|\psi^+\rangle_{aA_1}|\psi^+\rangle_{bB_2}(a_0 b_0 |00\rangle + a_0 b_1 |01\rangle + a_1 b_0 |10\rangle - a_1 b_1 |11\rangle)_{B_1 A_2} \\
&+ |\psi^+\rangle_{aA_1}|\psi^-\rangle_{bB_2}(a_0 b_0 |00\rangle - a_0 b_1 |01\rangle + a_1 b_0 |10\rangle + a_1 b_1 |11\rangle)_{B_1 A_2} \\
&+ |\psi^-\rangle_{aA_1}|\psi^+\rangle_{bB_2}(a_0 b_0 |00\rangle + a_0 b_1 |01\rangle - a_1 b_0 |10\rangle + a_1 b_1 |11\rangle)_{B_1 A_2} \\
&+ |\psi^-\rangle_{aA_1}|\psi^-\rangle_{bB_2}(a_0 b_0 |00\rangle - a_0 b_1 |01\rangle - a_1 b_0 |10\rangle - a_1 b_1 |11\rangle)_{B_1 A_2} \\
&+ |\psi^+\rangle_{aA_1}|\phi^+\rangle_{bB_2}(a_0 b_0 |01\rangle + a_0 b_1 |00\rangle - a_1 b_0 |11\rangle + a_1 b_1 |10\rangle)_{B_1 A_2} \\
&+ |\psi^+\rangle_{aA_1}|\phi^-\rangle_{bB_2}(a_0 b_0 |01\rangle - a_0 b_1 |00\rangle - a_1 b_0 |11\rangle - a_1 b_1 |10\rangle)_{B_1 A_2} \\
&+ |\psi^-\rangle_{aA_1}|\phi^+\rangle_{bB_2}(a_0 b_0 |01\rangle + a_0 b_1 |00\rangle + a_1 b_0 |11\rangle - a_1 b_1 |10\rangle)_{B_1 A_2} \\
&+ |\psi^-\rangle_{aA_1}|\phi^-\rangle_{bB_2}(a_0 b_0 |01\rangle - a_0 b_1 |00\rangle + a_1 b_0 |11\rangle + a_1 b_1 |10\rangle)_{B_1 A_2} \\
&+ |\phi^+\rangle_{aA_1}|\psi^+\rangle_{bB_2}(a_0 b_0 |10\rangle - a_0 b_1 |11\rangle + a_1 b_0 |00\rangle + a_1 b_1 |01\rangle)_{B_1 A_2} \\
&+ |\phi^+\rangle_{aA_1}|\psi^-\rangle_{bB_2}(a_0 b_0 |10\rangle + a_0 b_1 |11\rangle + a_1 b_0 |00\rangle - a_1 b_1 |01\rangle)_{B_1 A_2} \\
&+ |\phi^-\rangle_{aA_1}|\psi^+\rangle_{bB_2}(a_0 b_0 |10\rangle - a_0 b_1 |11\rangle - a_1 b_0 |00\rangle - a_1 b_1 |01\rangle)_{B_1 A_2} \\
&+ |\phi^-\rangle_{aA_1}|\psi^-\rangle_{bB_2}(a_0 b_0 |10\rangle + a_0 b_1 |11\rangle - a_1 b_0 |00\rangle + a_1 b_1 |01\rangle)_{B_1 A_2} \\
&+ |\phi^+\rangle_{aA_1}|\phi^+\rangle_{bB_2}(-a_0 b_0 |11\rangle + a_0 b_1 |10\rangle + a_1 b_0 |01\rangle + a_1 b_1 |00\rangle)_{B_1 A_2} \\
&+ |\phi^+\rangle_{aA_1}|\phi^-\rangle_{bB_2}(-a_0 b_0 |11\rangle - a_0 b_1 |10\rangle + a_1 b_0 |01\rangle - a_1 b_1 |00\rangle)_{B_1 A_2} \\
&+ |\phi^-\rangle_{aA_1}|\phi^+\rangle_{bB_2}(-a_0 b_0 |11\rangle + a_0 b_1 |10\rangle - a_1 b_0 |01\rangle - a_1 b_1 |00\rangle)_{B_1 A_2} \\
&+ |\phi^-\rangle_{aA_1}|\phi^-\rangle_{bB_2}(-a_0 b_0 |11\rangle - a_0 b_1 |10\rangle - a_1 b_0 |01\rangle + a_1 b_1 |00\rangle)_{B_1 A_2}]
\end{aligned} \qquad (4)$$

Here $|\psi^\pm\rangle = \frac{1}{\sqrt{2}}[|00\rangle \pm |11\rangle]$ and $|\phi^\pm\rangle = \frac{1}{\sqrt{2}}[|01\rangle \pm |10\rangle]$ are four Bell-states.

Alice and Bob perform Bell-state measurement (BSM) on their qubit pairs ($a, A_1$) and ($b, B_2$) respectively and convey their measurements result to each other through classical channel. After BSM, the collapsed states of qubits ($B_1, A_2$) are

$$|\mu^1\rangle_{B_1 A_2} = (a_0 b_0 |00\rangle + a_0 b_1 |01\rangle + a_1 b_0 |10\rangle - a_1 b_1 |11\rangle)_{B_1 A_2}$$

$$|\mu^2\rangle_{B_1 A_2} = (a_0 b_0 |00\rangle - a_0 b_1 |01\rangle + a_1 b_0 |10\rangle + a_1 b_1 |11\rangle)_{B_1 A_2}$$

$$|\mu^3\rangle_{B_1 A_2} = (a_0 b_0 |00\rangle + a_0 b_1 |01\rangle - a_1 b_0 |10\rangle + a_1 b_1 |11\rangle)_{B_1 A_2}$$

$$|\mu^4\rangle_{B_1A_2} = (a_0b_0|00\rangle - a_0b_1|01\rangle - a_1b_0|10\rangle - a_1b_1|11\rangle)_{B_1A_2}$$

$$|\mu^5\rangle_{B_1A_2} = (a_0b_0|01\rangle + a_0b_1|00\rangle - a_1b_0|11\rangle + a_1b_1|10\rangle)_{B_1A_2}$$

$$|\mu^6\rangle_{B_1A_2} = (a_0b_0|01\rangle - a_0b_1|00\rangle - a_1b_0|11\rangle - a_1b_1|10\rangle)_{B_1A_2}$$

$$|\mu^7\rangle_{B_1A_2} = (a_0b_0|01\rangle + a_0b_1|00\rangle + a_1b_0|11\rangle - a_1b_1|10\rangle)_{B_1A_2}$$

$$|\mu^8\rangle_{B_1A_2} = (a_0b_0|01\rangle - a_0b_1|00\rangle + a_1b_0|11\rangle + a_1b_1|10\rangle)_{B_1A_2}$$

$$|\mu^9\rangle_{B_1A_2} = (a_0b_0|10\rangle - a_0b_1|11\rangle + a_1b_0|00\rangle + a_1b_1|01\rangle)_{B_1A_2}$$

$$|\mu^{10}\rangle_{B_1A_2} = (a_0b_0|10\rangle + a_0b_1|11\rangle + a_1b_0|00\rangle - a_1b_1|01\rangle)_{B_1A_2}$$

$$|\mu^{11}\rangle_{B_1A_2} = (a_0b_0|10\rangle - a_0b_1|11\rangle - a_1b_0|00\rangle - a_1b_1|01\rangle)_{B_1A_2}$$

$$|\mu^{12}\rangle_{B_1A_2} = (a_0b_0|10\rangle + a_0b_1|11\rangle - a_1b_0|00\rangle + a_1b_1|01\rangle)_{B_1A_2}$$

$$|\mu^{13}\rangle_{B_1A_2} = (-a_0b_0|11\rangle + a_0b_1|10\rangle + a_1b_0|01\rangle + a_1b_1|00\rangle)_{B_1A_2}$$

$$|\mu^{14}\rangle_{B_1A_2} = (-a_0b_0|11\rangle - a_0b_1|10\rangle + a_1b_0|01\rangle - a_1b_1|00\rangle)_{B_1A_2}$$

$$|\mu^{15}\rangle_{B_1A_2} = (-a_0b_0|11\rangle + a_0b_1|10\rangle - a_1b_0|01\rangle - a_1b_1|00\rangle)_{B_1A_2}$$

$$|\mu^{16}\rangle_{B_1A_2} = (-a_0b_0|11\rangle - a_0b_1|10\rangle - a_1b_0|01\rangle + a_1b_1|00\rangle)_{B_1A_2} \quad (5)$$

Now, a nonlocal controlled phase gate operation is required to realize successful BQT. The controlled phase gate operation is described as $|00\rangle \to |00\rangle$, $|01\rangle \to |01\rangle$, $|10\rangle \to |10\rangle$ and $|11\rangle \to -|11\rangle$ in which the first qubit is the control qubit and the second qubit is the target qubit. Alice and Bob co-operate and apply quantum controlled phase gate operation (QCPGO) on qubits $A_2$ & $B_1$ with qubit $A_2$ as control qubit and qubit $B_1$ as target qubit.

**Table 1**: The collapsed state of qubits $B_1$ & $A_2$ corresponding to each BSM results of Alice and Bob and their corresponding unitary operations (UO).

| Alice's BSM | Bob's BSM | Collapsed state of qubits $B_1$ & $A_2$ | Alice's state after QCPGO | Bob's state after QCPGO | Alice's UO | Bob's UO |
|---|---|---|---|---|---|---|
| $|\psi^+\rangle_{aA_1}$ | $|\psi^+\rangle_{bB_2}$ | $|\mu^1\rangle_{B_1A_2}$ | $b_0|0\rangle + b_1|1\rangle$ | $a_0|0\rangle + a_1|1\rangle$ | $I$ | $I$ |
| $|\psi^+\rangle_{aA_1}$ | $|\psi^-\rangle_{bB_2}$ | $|\mu^2\rangle_{B_1A_2}$ | $b_0|0\rangle - b_1|1\rangle$ | $a_0|0\rangle + a_1|1\rangle$ | $\sigma_z$ | $I$ |
| $|\psi^-\rangle_{aA_1}$ | $|\psi^+\rangle_{bB_2}$ | $|\mu^3\rangle_{B_1A_2}$ | $b_0|0\rangle + b_1|1\rangle$ | $a_0|0\rangle - a_1|1\rangle$ | $I$ | $\sigma_z$ |
| $|\psi^-\rangle_{aA_1}$ | $|\psi^-\rangle_{bB_2}$ | $|\mu^4\rangle_{B_1A_2}$ | $b_0|0\rangle - b_1|1\rangle$ | $a_0|0\rangle - a_1|1\rangle$ | $\sigma_z$ | $\sigma_z$ |
| $|\psi^+\rangle_{aA_1}$ | $|\phi^+\rangle_{bB_2}$ | $|\mu^5\rangle_{B_1A_2}$ | $b_0|1\rangle + b_1|0\rangle$ | $a_0|0\rangle + a_1|1\rangle$ | $\sigma_x$ | $I$ |
| $|\psi^+\rangle_{aA_1}$ | $|\phi^-\rangle_{bB_2}$ | $|\mu^6\rangle_{B_1A_2}$ | $b_0|1\rangle - b_1|0\rangle$ | $a_0|0\rangle + a_1|1\rangle$ | $\sigma_z\sigma_x$ | $I$ |
| $|\psi^-\rangle_{aA_1}$ | $|\phi^+\rangle_{bB_2}$ | $|\mu^7\rangle_{B_1A_2}$ | $b_0|1\rangle + b_1|0\rangle$ | $a_0|0\rangle - a_1|1\rangle$ | $\sigma_x$ | $\sigma_z$ |
| $|\psi^-\rangle_{aA_1}$ | $|\phi^-\rangle_{bB_2}$ | $|\mu^8\rangle_{B_1A_2}$ | $b_0|1\rangle - b_1|0\rangle$ | $a_0|0\rangle - a_1|1\rangle$ | $\sigma_z\sigma_x$ | $\sigma_z$ |
| $|\phi^+\rangle_{aA_1}$ | $|\psi^+\rangle_{bB_2}$ | $|\mu^9\rangle_{B_1A_2}$ | $b_0|0\rangle + b_1|1\rangle$ | $a_0|1\rangle + a_1|0\rangle$ | $I$ | $\sigma_x$ |
| $|\phi^+\rangle_{aA_1}$ | $|\psi^-\rangle_{bB_2}$ | $|\mu^{10}\rangle_{B_1A_2}$ | $b_0|0\rangle - b_1|1\rangle$ | $a_0|1\rangle + a_1|0\rangle$ | $\sigma_z$ | $\sigma_x$ |
| $|\phi^-\rangle_{aA_1}$ | $|\psi^+\rangle_{bB_2}$ | $|\mu^{11}\rangle_{B_1A_2}$ | $b_0|0\rangle + b_1|1\rangle$ | $a_0|1\rangle - a_1|0\rangle$ | $I$ | $\sigma_z\sigma_x$ |
| $|\phi^-\rangle_{aA_1}$ | $|\psi^-\rangle_{bB_2}$ | $|\mu^{12}\rangle_{B_1A_2}$ | $b_0|0\rangle - b_1|1\rangle$ | $a_0|1\rangle - a_1|0\rangle$ | $\sigma_z$ | $\sigma_z\sigma_x$ |
| $|\phi^+\rangle_{aA_1}$ | $|\phi^+\rangle_{bB_2}$ | $|\mu^{13}\rangle_{B_1A_2}$ | $b_0|1\rangle + b_1|0\rangle$ | $a_0|1\rangle + a_1|0\rangle$ | $\sigma_x$ | $\sigma_x$ |

| | | | | | | |
|---|---|---|---|---|---|---|
| $\|\phi^+\rangle_{aA_1}$ | $\|\phi^-\rangle_{bB_2}$ | $\|\mu^{14}\rangle_{B_1 A_2}$ | $b_0\|1\rangle - b_1\|0\rangle$ | $a_0\|1\rangle + a_1\|0\rangle$ | $\sigma_z\sigma_x$ | $\sigma_x$ |
| $\|\phi^-\rangle_{aA_1}$ | $\|\phi^+\rangle_{bB_2}$ | $\|\mu^{15}\rangle_{B_1 A_2}$ | $b_0\|1\rangle + b_1\|0\rangle$ | $a_0\|1\rangle - a_1\|0\rangle$ | $\sigma_x$ | $\sigma_z\sigma_x$ |
| $\|\phi^-\rangle_{aA_1}$ | $\|\phi^-\rangle_{bB_2}$ | $\|\mu^{16}\rangle_{B_1 A_2}$ | $b_0\|1\rangle - b_1\|0\rangle$ | $a_0\|1\rangle - a_1\|0\rangle$ | $\sigma_z\sigma_x$ | $\sigma_z\sigma_x$ |

After controlled phase gate operation, the states $\|\mu^k\rangle_{B_1 A_2}$ where $k = 1,2,3...16$, decompose into the separate states of qubits $A_2$ & $B_1$ in the form $U^i(b_0\|0\rangle + b_1\|1\rangle)_{A_2} \otimes U^j(a_0\|0\rangle + a_1\|1\rangle)_{B_1}$, where $i, j = 0,1,2,3$ and $U^0, U^1, U^2, U^3$ are unitary operations to be done by the receiver Alice (Bob) depending upon BSM results of sender Bob (Alice). Corresponding to each BSM results of Alice and Bob, the collapsed states and the respective unitary operations are given in Table-1. Thus, Alice transmits an arbitrary single-qubit unknown information state $|\psi\rangle = a_0\|0\rangle + a_1\|1\rangle$ to Bob and Bob transmits an arbitrary single-qubit unknown information state $|\phi\rangle = b_0\|0\rangle + b_1\|1\rangle$ to Alice successfully. In this way, both Alice and Bob mutually exchange quantum information to each other. If both participants do not co-operate with each other, then no one can reconstruct the information sent to them, and therefore the exchange of information is possible only when both participants are honest to each other.

### 2.2. *Mutual exchange of certain class of multi-qubit entangled states*:

Now let us consider that the unknown quantum states possessed by Alice and Bob are multi-qubit entangled states $|\psi_m\rangle$ and $|\phi_n\rangle$ of the form given by

$$|\psi_m\rangle = (a_0|000.....0\rangle + a_1|111.....1\rangle)_{\alpha_1\alpha_2\alpha_3...\alpha_m} \tag{6}$$

$$|\phi_n\rangle = (b_0|000.....0\rangle + b_1|111.....1\rangle)_{\beta_1\beta_2\beta_3......\beta_n} \tag{7}$$

Here, $|\psi_m\rangle$ and $|\phi_n\rangle$ are m-qubit and n-qubit entangled states respectively.

The unknown coefficients $a_0, a_1, b_0 \& b_1$ satisfy the normalization conditions $|a_0|^2 + |a_1|^2 = 1$ and $|b_0|^2 + |b_1|^2 = 1$. Let Alice wants to transmit m-qubit state $|\psi_m\rangle$ to Bob and at the same time Bob wants to transmit n-qubit state $|\phi_n\rangle$ to Alice. For this purpose, in order to use the four-qubit cluster state $|Q\rangle_{A_1 B_1 A_2 B_2}$ as the quantum channel, Alice performs (m-1) CNOT operations on her qubits $\alpha_2, \alpha_3, ........\alpha_m$ with qubit $\alpha_1$ as control qubit and remaining (m-1) qubits as target qubits. This transforms m-qubit state $|\psi_m\rangle$ as

$$|\psi_m\rangle \xrightarrow{CNOT-Operations} (a_0|0\rangle + a_1|1\rangle)_{\alpha_1}|00.....0\rangle_{\alpha_2\alpha_3.....\alpha_m} \tag{8}$$

and Bob performs (n-1) CNOT operations on his qubits $\beta_2, \beta_3, ........\beta_n$ with qubit $\beta_1$ as control qubit and remaining (n-1) qubits as target qubits. This transforms n-qubit state $|\phi_n\rangle$ as

$$|\phi_n\rangle \xrightarrow{CNOT-Operations} (b_0|0\rangle + b_1|1\rangle)_{\beta_1}|00.....0\rangle_{\beta_2\beta_3.....\beta_n} \tag{9}$$

Thus, the actual task of transmissions of m-qubit and n-qubit entangled state reduces to the transmissions of arbitrary single-qubit states $(a_0|0\rangle + a_1|1\rangle)_{\alpha_1}$ and $(b_0|0\rangle + b_1|1\rangle)_{\beta_1}$. Now, by using the scheme discussed above in Section 2.1, Alice and Bob mutually exchange these arbitrary single qubit states to each other. After this, Alice and Bob reincarnate the desired multi-qubit entangled $|\phi_n\rangle$ and $|\psi_m\rangle$ respectively as follows:

(i) If m > n, then

- Bob introduces (m-n) auxiliary qubits in the state $|00......0\rangle_{X_1 X_2......X_{m-n}}$ and performs (m-1) CNOT operations with qubit-2 as control qubit and qubits $\beta_2, \beta_3,........\beta_n$, $X_1$, $X_2,.....,X_{m-n}$ as target qubits.
- Alice performs (n-1) CNOT operations with qubit-3 as control qubit and qubits $\alpha_2, \alpha_3,........\alpha_m$ as target qubits.

(ii) If m < n, then
- Alice introduces (n-m) auxiliary qubits in the state $|00......0\rangle_{Y_1 Y_2......Y_{n-m}}$ and performs (n-1) CNOT operations with qubit-3 as control qubit and qubits $\alpha_2, \alpha_3,........\alpha_m$, $Y_1$, $Y_2,.....,Y_{n-m}$ as target qubits.
- Bob performs (m-1) CNOT operations with qubit-2 as control qubit and qubits $\beta_2, \beta_3,........\beta_n$ as target qubits.

(iii) If m = n = k (say), then, Alice and Bob perform (k-1) CNOT operations on their qubits.
- Alice performs (k-1) CNOT operations with qubit-3 as control qubit and qubits $\alpha_2, \alpha_3,........\alpha_k$ as target qubits.
- Bob performs (k-1) CNOT operations with qubit-2 as control qubits and qubits $\beta_2, \beta_3,........\beta_k$ as target qubits.

Thus, Alice and Bob reincarnate the multi-qubit states $|\phi_n\rangle$ and $|\psi_m\rangle$ respectively and the mutual transmissions of multi-qubit entangled states (m↔n) is realized successfully.

## 3. Comparison

The comparison between the previous BQT schemes and the proposed BQT scheme has been made in Table-2 based upon the following aspects namely the unknown quantum states to be transmitted bi-directionally (QIT), the entangled state (ES) used as the quantum channel and the necessary non-local quantum operations (NNQO).

**Table 2**: The comparison between the previous BQT schemes and the proposed BQT scheme. Here, QCPG and CNOT respectively stand for quantum controlled phase gate and quantum controlled NOT gate.

| BQT schemes | QIT | Quantum Channel | NNQO | Inference |
|---|---|---|---|---|
| Ref. 28 | 1↔1 | 5-qubit ES | 1 non-local QCPG & 1 non-local CNOT | Successful |
| Ref. 29 | 1↔2 | 5-qubit ES | 2 non-local CNOT | Successful |
| Ref. 30 | 1↔1 | 6-qubit ES | 1 non local QCPG | Successful |
| Ref. 31 | 2↔2 | 4-qubit ES | Description of ES is incorrect | Failed |
| Ref. 33 | n↔n | 4-qubit ES | Description of ES is incorrect | Failed |
| Our scheme | 1↔1 | 4-qubit ES | 1 non local QCPG | Successful |
| Our generalized Scheme | m↔n | 4-qubit ES | 1 non local QCPG | Successful |

From Table 2, it is clear that in contrast to the previous BQT schemes,[28-30] the present BQ scheme requires reduced number of qubits in the quantum channel i.e., less quantum resource consumption and therefore it possesses higher intrinsic efficiency[33] and low quantum cost.

## 4. Conclusions

As four qubit cluster state has remarkable applications in one way quantum communications,[36-38] therefore, to find the use of four qubit cluster state in two way quantum communication is also remarkable. By utilizing four-qubit cluster state as the quantum channel, the authors[31, 33] tried to design a scheme for mutual exchange of quantum information between two legitimate participants. However, it has been found that in their scheme the transmission of quantum information cannot be achieved successfully. In the present study, we have shown that the four qubit cluster state shared between two legitimate participants can be used as a quantum channel for the mutual exchange of quantum information between them provided both participants cooperate and perform non local quantum controlled phase gate operation. We have also generalized the proposed BQT scheme to multi-qubit unknown quantum states. For teleporting certain class of multi-qubit states bi-directionally, first both participants perform local CNOT operations that reduce the original task into the BQT of arbitrary single qubit states. Finally, in order to reconstruct desired quantum states, both participants introduced desired number of auxiliary qubits and perform local CNOT operations. Eventually, the exchange of information is successfully possible only when both participants are honest to each other which is important in two way quantum communications.

**References**

1. C.H. Bennett, G. Brassard, C. Crépeau, R. Jozsa, A. Peres, and W.K. Wootters, *Phys. Rev. Lett.* **70** (1993) 1895.
2. G. Rigolin, *Phys. Rev. A, Gen. Phys.,* **71** (2005), Art. no.032303.
3. J. Dong and J. F. Teng, *Eur. Phys. J. D,* **49** (2008) 129.
4. P. Espoukeh and P. Pedram, *Quantum Inf. Process.,* **13**, (2014) 1789.
5. B. Zhang, X.-T. Liu, J. Wang, and C.-J. Tang, *Int. J. Theor. Phys.,* **55** (2016) 1601.
6. K. Nandi, and C. Mazumdar, *Int. J. Theor. Phys.,* **53** (2014) 1322.
7. X. Zuo, Y. Liu, W. Zhang, and Z. Zhang, *Sci. China, G, Phys., Mech. Astron.,* **52** (2009) 1906.
8. H.-J. Cao and H.-S. Song, *Int. J. Theor. Phys.,* **46** (2007) 1636.
9. Z.-X. Man, Y.-J. Xia, and N. B. An, *JETP Lett.,* **85** (2007) 662.
10. Z.-M. Liu and L. Zhou, *Int. J. Theor. Phys.,* **53** (2014) 4079.
11. Y.-H. Li, X.-L. Li, L.-P. Nie, and M.-H. Sang, *Int. J. Theor. Phys.,* **55** (2016) 1820.
12. M. Sisodia, and A. Pathak, *Int. J. Theor. Phys.,* **57** (2018) 2213.
13. M. Sisodia, A. Shukla, K. Thapliyal, and A. Pathak, *Quantum Inf. Process.,* **16** (2017) 1.
14. W.-M. Guo, and L.-R. Qin, *Chin. Phys. B,* **27** (2018), Art. no.110302.
15. X.-W. Zha, N. Miao, and H.-F. Wang, *Int. J. Theory Phys.,* **58** (2019) 2428.
16. N. Hao, Z.-H. Li, H.-Y. Bai, and C.-M. Bai, *Int. J. Theory Phys.,* **58** (2019) 1249.
17. W.-F. Cao, and Y.-G. Yang, *Int. J. Theory Phys.,* **58** (2019) 1202.
18. J. Wei, H.-Y. Dai, L. Shi, S. Zhao, and M. Zhang, *Int. J. Theory Phys.,* **57** (2018) 3104.
19. K. Hou, D.-Q. Bao, C.-J. Zhu, and Y.-P. Yang, *Quantum Inf. Process.,* **18** (2019) 104.
20. Y. Long, and Z. Shao, *Sci. Sin.-Phys. Mech. Astron.,* **49** (2019) Art. no.070301
21. Y.-H. Li, Y. Qiao, M.-H. Sang, and Y.-Y. Nie, *Int. J. Theor. Phys.,* **58** (2019) 2228.
22. Y.-R.Sun, G. Xu, X.-B. Chen, Y. Yang, and Y.-X. Yang, *IEEE Access,* **7** (2019) 2811.
23. R.-G. Zhou, R. Xu, and H. Lan, *IEEE Access,* **7** (2019) 44269.
24. P.-C. Ma, G.-B. Chen, X.-W. Li, and Y.-B. Zhan, *Int. J. Theor. Phys.,* **57** (2018) 443.
25. J. Shi, P.-C. Ma, and G.-B. Chen, *Int. J. Theor. Phys.,* **58**, (2019) 372.
26. R.-G. Zhou, C. Qian, and H. Ian, *IEEE Access*, **7**, (2019) 4244.
27. X.W. Zha, Z.C. Zou, J.X. Qi, and H.Y Song, *Int. J. Theor. Phys.,* **52** (2013) 1740.
28. Y. Chen, *Int. J. Theor. Phys.,* **53** (2014) 1454.
29. M. Sang, *Int. J. Theor. Phys.,* **55** (2016) 1333.
30. A .Yan, *Int. J. Theor. Phys.,* **52** (2013) 3870.
31. R. Zhou, C. Qian and H. Ian, *Int. J. Theor. Phys.,* **58** (2019)150.
32. M.S.S. Zadeh, M. Houshmand, and H. Aghababa, *Int. J. Theor. Phys.,* **56** (2017) 2101.
33. P. Kazemikhah and H. Aghababa, *Int. J. Theor. Phys.,* **60** (2021) 378.
34. V. Verma, *Int. J. Theor. Phys.,* **59** (2020) 3329.

35. H.J. Briegel and R. Raussendorf, *Phys. Rev. Lett.* **86** (2001) 910.
36. L. S. Song, Y.-Y. Nie, Z.-H. Hong, X.-J. Yi, and Y.-B. Huang, *Commun. Theor. Phys.* **50** (2008) 633.
37. S. Murlidharan and P. K. Panigrahi, *Phys. Rev. A,* **78** (2008) 062333.
38. Q. N. Zhang, C.-C. Li, Y.-H. Li, and Y.-Y. Nie, *Int. J. Theor. Phys.,* **52** (2013) 22.